\documentclass[aps,prb,showpacs,reprint,amsmath,amssymb,floatfix,superscriptaddress,]{revtex4-2}

\usepackage{hyperref}
\usepackage{graphicx}
\usepackage{bm}
\usepackage{dsfont}
\usepackage{xcolor}
\usepackage[utf8]{inputenc}
\usepackage[normalem]{ulem}
\usepackage{physics}

\newcommand{\beq}{\begin{equation}}
\newcommand{\eeq}{\end{equation}}
\newcommand{\beqa}{\begin{eqnarray}}
\newcommand{\eeqa}{\end{eqnarray}}

\begin{document}

\title{Absence of Majorana oscillations in finite-length full-shell hybrid nanowires}
\author{Carlos Payá}
\affiliation{Instituto de Ciencia de Materiales de Madrid (ICMM), CSIC, Madrid, Spain}
\author{Pablo San-Jose}
\affiliation{Instituto de Ciencia de Materiales de Madrid (ICMM), CSIC, Madrid, Spain}
\author{Carlos J. Sánchez Martínez}
\affiliation{Departamento de Física Teórica de la Materia Condensada, Universidad Autónoma de Madrid, Madrid, Spain}
\affiliation{Condensed Matter Physics Center (IFIMAC), Universidad Autónoma de Madrid, Madrid, Spain.}
\author{Ramón Aguado}
\affiliation{Instituto de Ciencia de Materiales de Madrid (ICMM), CSIC, Madrid, Spain}
\author{Elsa Prada}
\email{elsa.prada@csic.es}
\affiliation{Instituto de Ciencia de Materiales de Madrid (ICMM), CSIC, Madrid, Spain}

\date{\today}

\begin{abstract}
Majorana bound states (MBSs) located at the ends of a hybrid superconductor-semiconductor  nanowire are only true zero modes if their characteristic localization length is much smaller than the nanowire length, $\xi_M\ll L$. Otherwise, their wave function overlap gives rise to a  finite energy splitting that shows a characteristic oscillatory pattern $\sim e^{-2L/\xi_M}\cos(k_F L)$ versus external parameters that modify the Fermi momentum $k_F$. Detecting such ``Majorana oscillations", measurable through low-bias conductance, has been proposed as a strategy for Majorana detection in pristine nanowires. Here we discuss how this detection scheme does not work in full-shell hybrid nanowires, an alternative design to partial-shell nanowires in which a superconductor shell fully wraps the semiconductor core. Using microscopic models, we provide both numerical simulations for Al/InAs hybrids as well as analytical approximations in terms of general nanowire parameters. We find that Majorana oscillations with flux in full-shell nanowires are absent in a wide portion of parameter space. This absence is not a signature of non-overlapping left- and right-end MBSs, but a consequence of the Majorana oscillation period being systematically larger than the flux window of odd Little-Parks lobes where Majorana zero-energy peaks are predicted to appear. Our results demonstrate that split near-zero modes or individual zero-energy crossings should not be dismissed as trivial even if they are found not to oscillate with flux.
\end{abstract}


\maketitle

\section{Introduction}

\begin{figure}[h]
   \centering
   \includegraphics[width=\columnwidth]{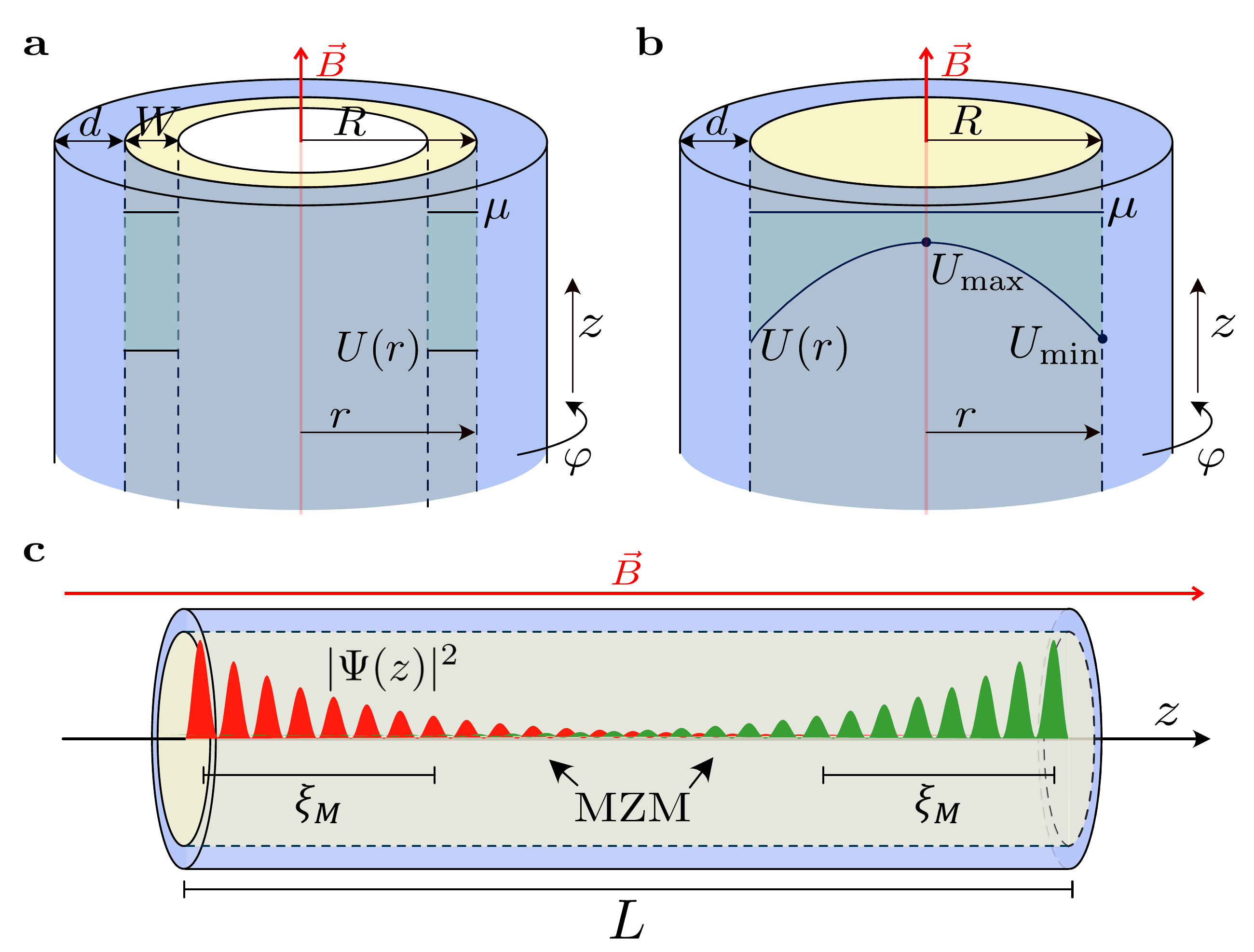}
   \caption{\textbf{Full-shell hybrid nanwowire.} (a) Sketch of a full-shell nanowire with a tubular-core geometry. An insulating core (white) is surrounded by a semiconductor tube (yellow) of external radius $R$ and thickness $W$, and is completely encapsulated in a thin superconductor shell (blue) of thickness $d$. In an applied axial magnetic field $B$ the hybrid wire is threaded by a non-quantized magnetic flux $\Phi=\pi(R+d/2)^2B$. The chemical potential $\mu$ and the radial electrostatic potential energy $U(r)$ are schematically depicted inside. (b) Same as (a) but for a semiconductor solid-core geometry. The conduction-band bottom inside the semiconductor exhibits a dome-like radial profile with maximum value at the center, $U_{\rm{max}}$, and minimum value at the superconductor-semiconductor interface, $U_{\rm{min}}$. (c) Sketch of a full-shell nanowire of finite length $L$ in the topological phase. In red and green, Majorana bound state (MBS) wavefunctions at the left and right ends of the nanowire along the longitudinal direction. $\xi_{\rm M}$ is the Majorana localization length (defined in a semi-infinite wire).}
   \label{fig:sketch}
\end{figure}

Majorana bound states (MBSs) are non-Abelian quasiparticles that appear at the ends of one-dimensional topological superconductors \cite{Kitaev:AoP03,Das-Sarma:NQI15,Nayak:RMP08,Aguado:PT20}. Among the various proposed platforms that host MBSs \cite{Yazdani:S23, Alicea:RPP12, Beenakker:SPLN20, Aguado:RNC17}, arguably the most studied one is known as the Oreg-Lutchyn Majorana nanowire \cite{Lutchyn:PRL10,Oreg:PRL10}. It is based on hybrid superconductor-semiconductor heterostructures with strong spin-orbit coupling (SOC), which may undergo a topological transition in response to an applied Zeeman field.
During the last two decades, several measurement schemes of increasing complexity have been pursued to unambiguously confirm the emergence of MBSs, including tunneling~\cite{Mourik:S12,Vaitiekenas:NP21}, Coulomb~\cite{Vaitiekenas:S20,Valentini:N22} and quantum-dot-assisted spectroscopies~\cite{Deng:PRB18}. More recently, sophisticated protocols~\cite{MicrosoftQuantum:PRB23} that combine local and non-local conductance measurements using time-resolved interferometry and single-shot parity readout~\cite{MicrosoftQuantum:24} have also been implemented. Despite such experimental efforts, interpreting the data is challenging when taking into account all the phenomenology \cite{Prada:NRP20} that arises in realistic hybrid systems beyond the predictions of the ideal single-mode Majorana nanowire model. Generalizations include e.g. the effects of disorder \cite{Pan:PRR20,Das-Sarma:PRB21,Das-Sarma:NP23} and material inhomogeneities~\cite{Kells:PRB12,Prada:PRB12,Rainis:PRB13,Roy:PRB13,Liu:PRB17,Penaranda:PRB18,Vuik:SP19,Rossi:PRB20}.

In pristine Oreg-Lutchyn nanowires, a way to identify MBSs was proposed already in 2012~\cite{Das-Sarma:PRB12}, see also Refs. \onlinecite{Klinovaja:PRB12,Lim:PRB12}. The effective size of the MBS wavefunction is called the Majorana localization length, $\xi_M$, defined in a semi-infinite wire. In a finite-length nanowire whose length $L$ is not much larger than $\xi_M$, the left- and right-end MBS wave functions spatially overlap, see Fig. \ref{fig:sketch}. This gives rise to an energy splitting $\sim e^{-2L/\xi_M}\cos(k_F L)$ of the zero-energy Majorana mode that decays exponentially with $L$, just as the overlap \footnote{Note that there exist different definitions of the left and right Majorana wave function overlap, see Ref. \onlinecite{Penaranda:PRB18}. However, they all decay exponentially as $e^{-2L/\xi_M}$.}, but that also has a sinusoidal behavior with Fermi momentum $k_F$. This in turn could be detected as a characteristic oscillatory splitting with Zeeman field (or chemical potential) of the zero-bias conductance peak in tunneling spectroscopy measurements.

While Majorana oscillations have been studied in different contexts in Zeeman-driven nanowires~\cite{Fleckenstein:PRB18,Cao:PRL19}, including critical current oscillations in Josephson junctions~\cite{Cayao:PRB17}, they have not yet been analyzed in a full-shell hybrid geometry. Full-shell hybrid nanowires are an alternative Majorana nanowire design that came to the spotlight in 2020~\cite{Vaitiekenas:S20} in which the superconductor shell that induces superconductivity into the semiconductor core is not limited to some facets of the nanowire, but wraps it all around. The doubly-connected geometry of the superconductor has profound implications in the response of these wires to an applied axial magnetic field. It gives rise for instance to the Little-Parks (LP) effect~\cite{Little:PRL62, Parks:PR64}, whereby the superconducting order parameter oscillates with flux in a series of \emph{lobes} characterized by an integer number $n$ of superconductor phase windings, called fluxoids. It also enables the presence of a special type of subgap Andreev states \cite{Tinkham:96, Caroli:PL64, BrunHansen:PLA68, Kopasov:PSS20, Kopasov:PRB20} that have been called Caroli-de Gennes-Matricon (CdGM) analog states \cite{San-Jose:PRB23}. In the topological phase, one of those subgap states transforms into a Majorana zero mode, extending across a flux interval inside odd-$n$ LP lobes \cite{Vaitiekenas:S20,Penaranda:PRR20,Paya:PRB24}. Importantly, the topological superconducting properties are not driven by the Zeeman effect, but by the orbital effect of the magnetic field, which among other things permits operating these wires at much smaller magnetic field that their partial-shell counterparts. The rich phenomenology of these wires is being analyzed in recent years~\cite{Woods:PRB19,Vaitiekenas:PRB20,Valentini:S21,Vekris:SR21,Escribano:PRB22,Valentini:N22,Razmadze:PRB24,Giavaras:PRB24,Klausen:N23,Ibabe:NC23,Paya:PRB24,Ibabe:NL24,Valentini:24}, but there are still many basic questions that need to be understood.

In this work we consider the effect a of finite nanowire length on the Majorana zero modes. We find that the expected Majorana oscillations are absent for a wide (and realistic) region of parameter space. This hinders the possibility of using Majorana oscillations as a way to identify and characterize MBSs in full-shell wires. We consider both microscopic numerical simulations in perfect Al/InAs full-shell nanowires as well as general analytical derivations as a function of the hybrid parameters. The absence of oscillations is not a consequence of the left and right-end MBSs not overlapping in a finite-$L$ nanowire. Actually, zero energy modes in full-shell geometries still split as $\sim e^{-2L/\xi_M}\cos(k_F L)$ when the nanowire length $L$ is not much larger than the Majorana localization length $\xi_M$. The reason is that, except for very narrow and long wires, the Majorana oscillation period with flux, which we show to be $\delta\Phi=2\pi/|L\partial_\Phi k_F(\Phi)|$, is systematically of the order or larger than the lobe flux window. The practical consequence is that a lack of oscillations around zero of a near-zero mode or parity crossing cannot be used to rule out a topological origin.

This paper is organized as follows. In Sec. \ref{Sec:results} we explain the two different realistic models that we consider for the full-shell hybrid geometry: the tubular-core nanowire, analyzed in Sec. \ref{Sec:TCNw}, and the solid core nanowire, analyzed in Sec. \ref{Sec:SCNw}. In Sec. \ref{Sec:discussion} we provide a discussion of the results. A summary of our findings and the conclusions of this work are given in Sec. \ref{Sec:conclusions}. In Appendix \ref{Ap:Hamiltonian} we provide the Hamiltonians employed to analyze these hybrid wires, including the tubular-core approximation and the mapping to the Oreg-Lutchyn model (Appendix \ref{Ap:TC}). An analytical expression for the Majorana oscillation period in full-shell hybrid nanowires is derived in Appendix \ref{Ap:oscillations}, together with an analysis of the typical range of parameters for which there are no Majorana oscillations. While the numerical results we present in the main text correspond to a wire in the non-destructive LP regime, the destructive one is considered in Appendix \ref{Ap:DestructiveLP}.

\section{Models and results}
\label{Sec:results}

A full-shell hybrid nanowire with a semiconductor core of radius $R$ and a thin superconductor shell of thickness $d$ is shown in Fig. \ref{fig:sketch}. We consider two possible realistic models for the semiconductor core \cite{Paya:PRB24}. In the first one, Fig. \ref{fig:sketch}(a), the interior is insulating around the axis and the semiconductor has a tubular shape of thickness $W$. This is called the tubular-core model. In the second one, Fig. \ref{fig:sketch}(b), the semiconductor completely fills the interior of the shell. This is dubbed the solid-core model.

In our models we employ cylindrical coordinates, with radial coordinate denoted by $r$, azimuthal angle by $\varphi$ and axial coordinate chosen along the $z$ direction. For simplicity, we assume that the hybrid wire has cylindrical symmetry. It has been shown that this is a very good approximation for a more realistic hexagonal-shaped nanowire \cite{San-Jose:PRB23,Paya:PRB24}. The full-shell nanowire is threaded by a magnetic field $\vec{B}=B\hat{z}$ that gives rise to a flux $\Phi = \pi R_\mathrm{LP}^2 B$, where we define the LP radius as $R_\mathrm{LP} =R + d/2$, the mean radius of the shell.

The methodology to analyze these wires, both analytically and numerically, has been thoroughly described in Refs. \cite{Vaitiekenas:S20,San-Jose:PRB23,Paya:PRB24}. We provide a summary in Appendix \ref{Ap:Hamiltonian}. As discussed there, it is possible to arrive at an effective Bogoliubov-de Gennes (BdG) Hamiltonian for the hybrid system, see Eq. \eqref{eq:solidrot}. The Hamiltonian is expressed in terms of the flux $\Phi$, which, in units of the superconducting flux quantum $\Phi_0=h/2e$ (where $h$ is the Plank constant and $e$ the electron charge), defines the fluxoid $n(\Phi) = \lfloor \Phi/\Phi_0\rceil$. This is the number of times the superconductor phase winds around the wire axis for a given flux. Other variables include nanowire intrinsic parameters such as the effective mass $m$, the chemical potential $\mu$, the SOC $\alpha$ and the $g-$factor, the geometrical parameters ($d$, $R$, $W$), the radial electrostatic potential energy $U(r)$, the decay rate from the semiconductor core into the superconductor $\Gamma_{\rm S}$, the shell superconducting gap at zero field $\Delta_0$, the diffusive superconducting coherence length $\xi_d$, and the radial and generalized angular momentum quantum numbers $(m_r,m_J)$ that label the different transverse subbands of the wire.

Previous results in Ref. \cite{Paya:PRB24} analyzed in detail the phenomenology of a semi-infinite full-shell nanowire, including the topological phase diagrams as a function of $R$, $W$, $\alpha$, $\mu$ and $\Gamma_{\rm S}$, the local density of states (LDOS) at the end of the wire, and the differential conductance through an normal-superconductor junction. In this work we consider the phenomenology of these wires when they have a finite length $L$ so that, in the topological regime, left- and right-end  MBSs can overlap, see Fig. \ref{fig:sketch}(c).


In our numerical simulations of the main text and the Appendix \ref{Ap:DestructiveLP} we will employ typical parameters for an Al/InAs hybrid nanowire, since those are the most common materials analyzed experimentally so far. We note that, recently, a practical theoretical proposal based on semiconductor hole bands (instead of electron bands as here) was studied in Ref. \cite{Vezzosi:24} for InP/GaSb core-shell nanowires. General analytical results for the Majorana splittings can be found in Appendix \ref{Ap:oscillations}.

\subsection{Tubular-core nanowire}
\label{Sec:TCNw}

\begin{figure*}
\centering
\includegraphics[width=\textwidth]{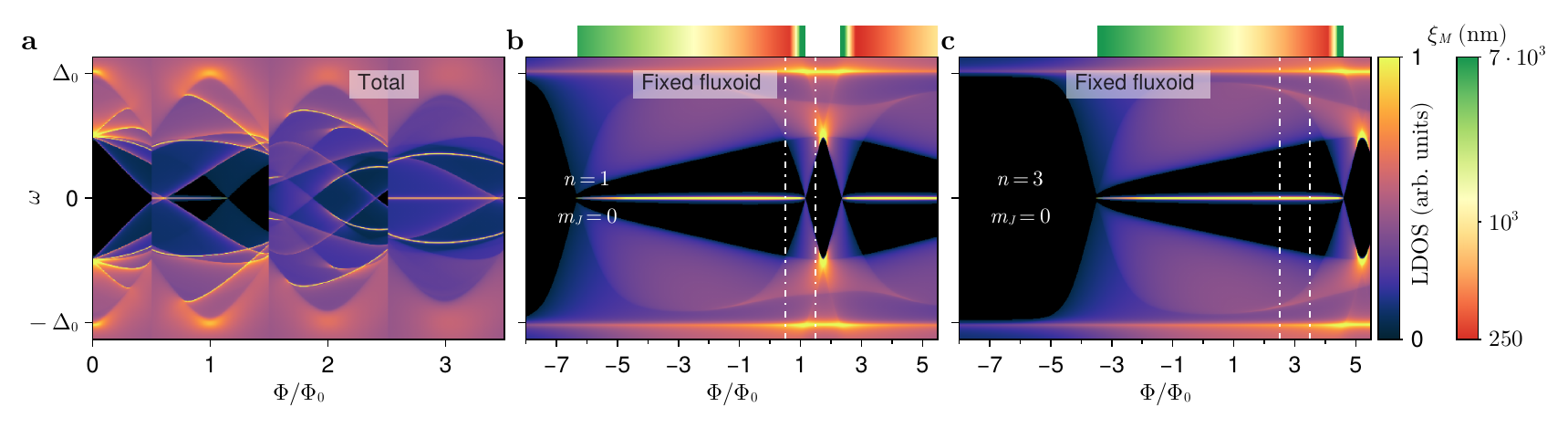}
\caption{\textbf{Tubular-core nanowire of semi-infinite length.} (a) Local density of states (LDOS) at the end of a semi-infinite $d=10$~nm, $R=70$~nm, and $W=20$~nm tubular-core nanowire (in arbitrary units) as a function of energy $\omega$ and applied normalized magnetic flux $\Phi/\Phi_0$, displaying half of the $n=0$, and the full $n=1,2,3$ Little-Parks (LP) lobes. Majorana zero modes are visible in the $n=1$ and $n=3$ LP lobes. The full-shell wire is in the non-destructive LP regime. Other subgap features are Caroli--de Gennes--Matricon (CdGM) analog states. (b) Contribution of the subbands with $m_J=0$ generalized-angular-momentum quantum number to the LDOS in (a) but suppressing the LP modulation and fixing the fluxoid number to $n=1$ for all the flux range displayed. The boundaries of the first LP lobe are marked with vertical dashed lines. The Majorana localization length $\xi_M$ (in nm) as a function of flux is shown above with a color bar. (c) Same as (b) but for the $n=3$ fluxoid number. Minimum $\xi_M$ inside lobe $n=1$: $\sim 260$~nm, inside lobe $n=3$: $\sim 290$~nm. Other parameters: $\alpha = 50$~meVnm,  $\mu = 18.1$~meV, $g = 10$,  $\Gamma_{\rm S} = 10 \Delta_0$, $\Delta_0 = 0.23$~meV, $\xi_d = 70$~nm.
}
\label{fig:TCMsemi}
\end{figure*}

We start by analyzing a full-shell hybrid nanowire with a tubular core, see Fig. \ref{fig:sketch}(a). This model could describe a multi-shell nanowire with an insulating core and a semiconductor shell (a so-called core-shell nanowire in the literature), fully wrapped in a superconductor shell.

In Fig. \ref{fig:TCMsemi}(a) we can see the LDOS of a semi-infinite tubular-core nanowire. It is represented against the flux $\Phi$ through the LP section normalized to  $\Phi_0$. This flux causes the superconducting phase in the shell to acquire a quantized winding $n = 0, \pm 1, \pm 2, \dots$ around the nanowire axis. Winding number jumps are accompanied by a repeated suppression and recovery of the superconductor shell gap, forming so-called LP \textit{lobes} as a function of flux, characterized by the integer number $n$ of fluxoids through the section. In Fig. \ref{fig:TCMsemi}(a) we show half of the $n=0$ and the full $n=1-3$ lobes. We observe a number of bright features below the parent gap in the different lobes. These are the so-called CdGM analog states analyzed in detail Ref.~\cite{San-Jose:PRB23,Paya:PRB24}. They are Van Hove singularities that are induced by the superconductor shell on the normal core bands. The geometrical parameters (core radius $R=70$~nm and shell thickness $d=10$~nm) and the superconductor parameters (gap at zero field $\Delta_0=0.23$~meV and diffusive superconducting coherence length $\xi_d=70$~nm) are such that the wire is in the non-destructive LP regime, meaning that the parent gap does not close between lobes at half integer values of $\Phi_0$. As a result, all the subgap features change abruptly at fluxoids jumps between lobes. The transition from the non-destructive to the destructive LP regime~\cite{Schwiete:PRB10}, analyzed in Appendix \ref{Ap:DestructiveLP}, typically happens for smaller $R_{\rm LP}$ radius, see Eq. \eqref{eq:desborder}.

In a full-shell nanowire that can be described with a cylindrical approximation, MBSs are predicted to appear at odd LP lobes when the parameters of the wire are such that the system is the topological phase, see Appendix \ref{Ap:Hamiltonian}. When the thickness of the tubular semiconductor, $W$, is small compared to its radius $R$, it is possible to approximate the radial electrostatic energy $U(r)$ and the SOC $\alpha(r)$ as constants, see Fig. \ref{fig:sketch}(a). When the doping of the semiconductor wire is small, as it is typically the case in depleted nanowires, only the lowest radial subband is populated and the Hamiltonian of the system can be approximated by Eq. \eqref{TCHam}, given in terms of the average radius $R_{\rm av}$ of electron wave functions inside the semiconductor. In this case, for the subband with lowest generalized angular momentum $m_J=0$ there is a direct mapping to the single-mode Oreg-Lutchyn model, see Ref. \cite{Vaitiekenas:S20} and Appendix \ref{Ap:TC}. It is then straightforward to understand the appearance of Majorana zero energy modes in odd lobes, since these are the only ones where $m_J$ can take a zero value. This mapping is also very useful to appreciate that the topological phase transition in full-shell nanowires is driven by an orbital effect, see the expression for the \textit{effective} Zeeman field, Eq. \eqref{LutchynZeeman}.

Since in Fig. \ref{fig:TCMsemi}(a) the values of $\alpha$, $\mu$ and $\Gamma_{\rm S}$ have been chosen so that the full-shell wire is in the topological regime, we can observe a zero-energy peak (ZEP) in the $n=1$ lobe that emerges due to the presence of a MBS at the end of the semi-infinite wire at zero energy. It extends from the left end of the lobe to an intermediate value of flux within the lobe, as corresponds to a tubular-core nanowire. At the right end of the ZEP, the topological band closing and reopening of the $m_J=0$ subband is visible. At the left end a CdGM state crosses zero energy. Still, there is a region of fluxes where the Majorana mode is spectrally separated from other subgap states and is thus protected by a topological minigap. For the parameters of this wire, there is also a ZEP in the $n=3$ lobe that crosses the whole lobe. However, it coexists with a finite LDOS background from the rest of the occupied CdGM analogs, so it is unprotected.

\begin{figure*}
\centering
\includegraphics[width=\textwidth]{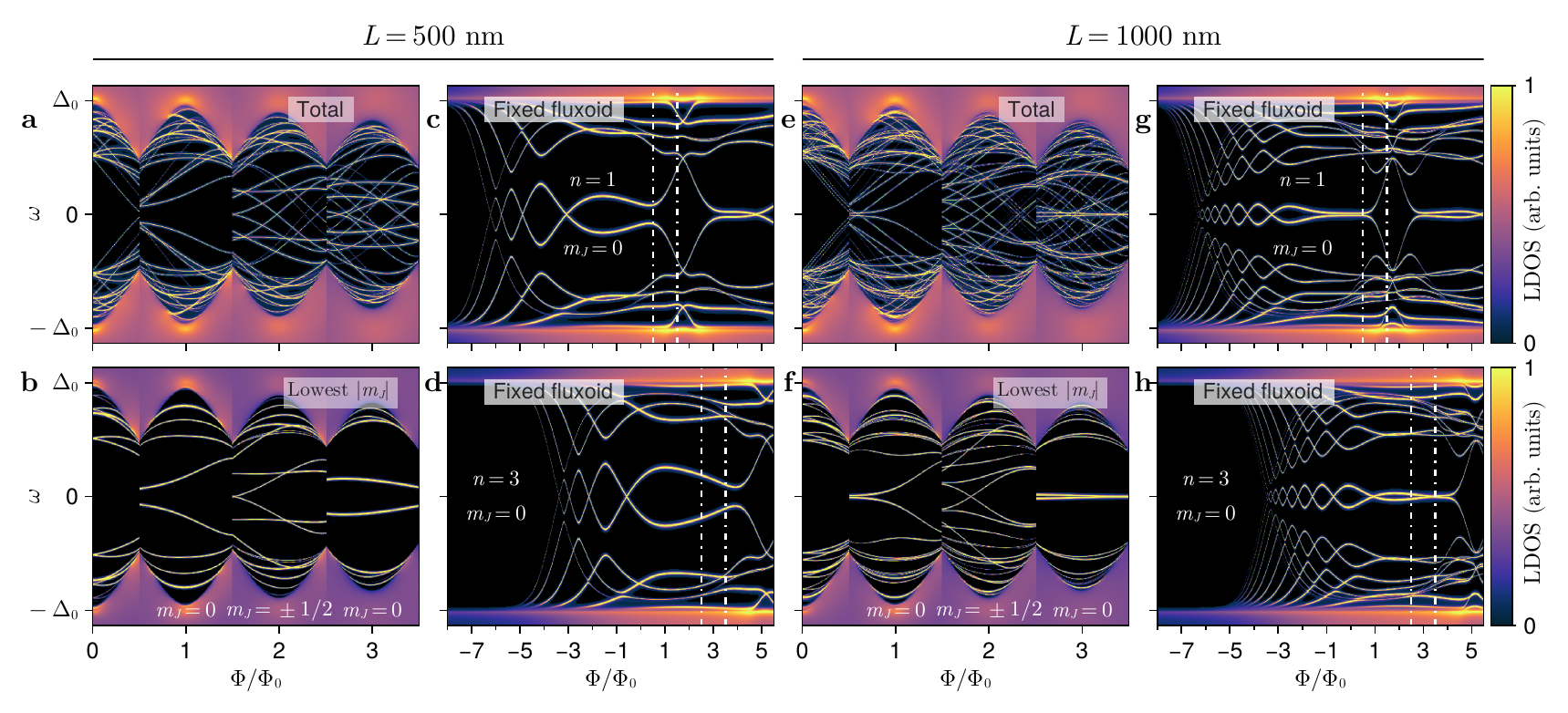}
\caption{\textbf{Tubular-core nanowire of finite length.} (a) Local density of states (LDOS) at the end of a $L=500$~nm, $d=10$~nm, $R=70$~nm, and $W=20$~nm tubular-core nanowire (in arbitrary units) as a function of energy $\omega$ and applied normalized flux $\Phi/\Phi_0$, displaying half of the $n=0$, and the full $n=1,2,3$ LP lobes. (b) Contribution of the lowest $m_J$ quantum numbers to the LDOS in (a) ($m_J=0$ for odd lobes, $m_J=\pm 1/2$ for even lobes). (c) $m_J=0$ contribution to the LDOS in (a) but suppressing the LP modulation and fixing the fluxoid number to $n=1$ for all the magnetic-flux range displayed. The boundaries of the first LP lobe are marked with vertical dashed lines. (d) Same as (c) but for the $n=3$ fluxoid number. (e-h) Same as (a-d) but for a $L=1000$~nm full-shell nanowire. Other parameters like in Fig. \ref{fig:TCMsemi}.
}
\label{fig:TCMfinite}
\end{figure*}

For illustrative purposes, in Figs. \ref{fig:TCMsemi}(b) and (c) we show the $m_J=0$ contribution to the LDOS as a function of flux by artificially fixing the fluxoid number to $n=1$ and $n=3$, respectively. We moreover remove the LP modulation [by taking $\xi_d=0$ in Eq. \eqref{depairing}] to observe more clearly the subgap phenomenology. Only the magnetic flux region bounded within vertical dashed lines (the actual lobe) corresponds to a stable real solution, but we display a much larger flux range to observe the whole interval of the Majorana zero-energy anomaly, from its appearance (sometimes within the actual lobe) to its disappearance (typically at large negative flux). One could think of this plot as a metastable solution of the full-shell wire if it were possible to quickly vary the magnetic flux so that the the system does not reach its ground state, thus leading to a fixed fluxoid. 
While this is probably possible for some flux range close to the actual lobe, it is clearly impossible for the magnetic range shown. Still, this kind of plots provide a useful view into the behavior of the Majorana oscillations with flux in a finite-length wire and their absence within the flux windows of odd lobes.

In Fig. \ref{fig:TCMsemi}(b) the ZEP within the $n=1$ lobe is dominated by the $k_z=0$ gap of the $m_J=0$ subband (the so-called inner gap that undergoes the topological inversion), whereas the one at $k_z=k_F$ (the outer gap) takes over for flux values outside the lobe. The Majorana localization length $\xi_M$ is an important quantity to later understand Majorana overlap in the finite length case. In Fig. \ref{fig:TCMsemi} we compute $\xi_M$ numerically~\cite{Paya:PRB24} and display it above the LDOS plots with a color bar (again, only the part within the vertical dashes lines is physically meaningful). In this case, $\xi_M$ reaches a minimum of $\xi_M\approx 260$~nm precisely at the left end of the $n=1$ lobe. $\xi_M$ is larger in the third lobe (with a minimum of $\approx 290$~nm) as the ZEP is dominated there by a smaller $m_J=0$ minigap, namely the outer gap, which monotonically decreases until it closes for negative $\Phi/\Phi_0$. The behaviour described above is typical for tubular-core models.

\begin{figure*}
\centering
\includegraphics[width=\textwidth]{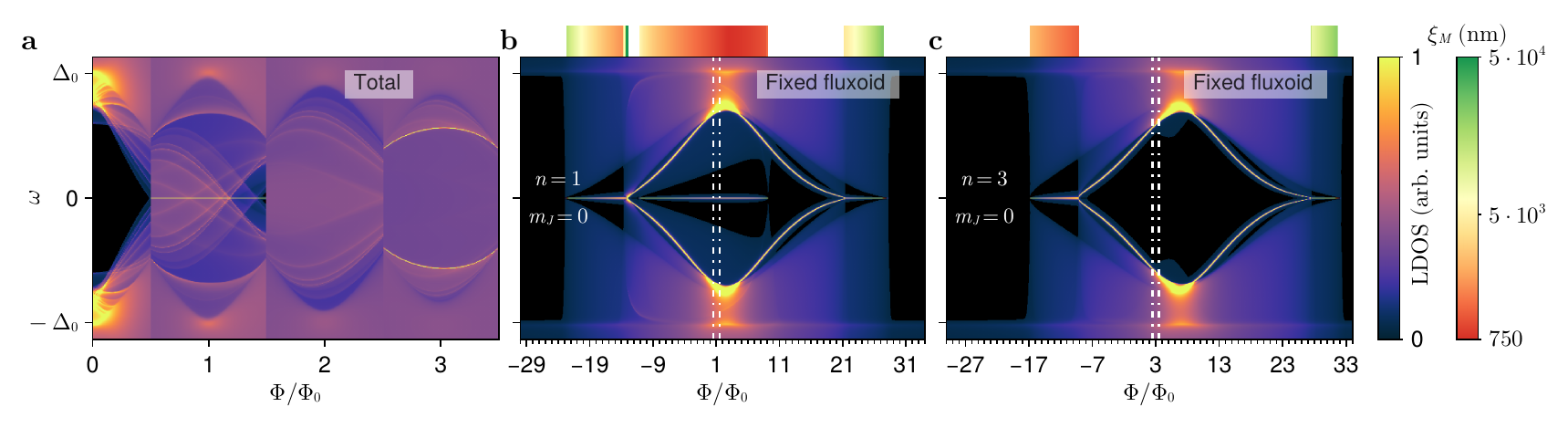}
\caption{\textbf{Solid-core nanowire of semi-infinite length.} Same as Fig. \ref{fig:TCMsemi} but for a solid-core nanowire with $d=10$~nm and $R=70$~nm. The radial dome-like electrostatic potential profile inside the semiconductor has $U_{\rm min}=-30$~meV and $U_{\rm max}=0$, with $\mu = 2$~meV. Other parameters like in Fig. \ref{fig:TCMsemi}, except for $\langle \alpha \rangle = 20$~meVnm and $\Gamma_{\rm S}= 40 \Delta_0$. Minimum $\xi_M$ inside lobe $n=1$: $\sim 890$~nm.
}
\label{fig:SCMsemi}
\end{figure*}

If now we consider the same parameters as in Fig. \ref{fig:TCMsemi} but for a finite-length nanowire, we arrive at the behavior observed in Fig. \ref{fig:TCMfinite}. The total LDOS for $L=500$~nm can bee seen in Fig. \ref{fig:TCMsemi}(a). This length is approximately twice the Majorana localization length within the first lobe, so that the left an right Majoranas overlap considerably. Two striking differences appear with respect to the semi-infinite case. On the one hand, the smooth Van Hove signals that characterized the CdGM analogs in Fig. \ref{fig:TCMsemi}(a) are now transformed into a series of discrete levels for each $m_J$. These levels are longitudinally confined CdGM states. On the other hand, the Majorana ZEP splits into two trivial states, with a splitting that grows as the length is reduced. To see this more clearly, in Fig. \ref{fig:TCMfinite}(b) we plot the contribution to the LDOS of only the lowest $|m_J|$ subbands for each lobe, i.e. $m_J=0$ ($m_J=\pm 1/2$) for odd (even) lobes. Notice that the third lobe ZEP is also split. The fixed-fluxoid $m_J=0$ LDOS plots for $n=1$ and $n=3$ are shown in Figs. \ref{fig:TCMfinite}(c,d), respectively. It is possible to see that for large negative fluxes the ZEP does indeed oscillate with magnetic field, but this happens far away from the measurable region within the lobe boundaries (vertical dashed white lines).

Corresponding results for an $L=1000$~nm wire are shown in Figs. \ref{fig:TCMfinite}(e-h). Since the wire is longer, the Majorana oscillations in Figs. \ref{fig:TCMfinite}(g,h) have smaller amplitude and shorter period, so that there is a larger number of oscillations along the full flux range of the zero-energy anomaly. Still, within the $n=1$ and $n=3$ lobes we just observe a split Majorana state, but not a full oscillation. The reason is twofold. First, at least in the $n=1$ lobe, we are accessing only the beginning of the Majorana zero mode after the topological phase transition, where the ZEP has not yet started to oscillate significantly because it is dominated by the $k_z=0$ gap, see Figs. \ref{fig:TCMfinite}(c,g). Second, and more importantly, the oscillation period of the Majorana is larger than the flux  range it occupies within the lobe. This can be seen for instance in Fig. \ref{fig:TCMfinite}(h).
One may ask whether the absence of Majorana oscillations is a peculiarity specific to tubular-core models but not to solid-core ones. The answer is negative.  We analyze this in the next section.

\subsection{Solid-core nanowire}
\label{Sec:SCNw}

\begin{figure*}
\centering
\includegraphics[width=\textwidth]{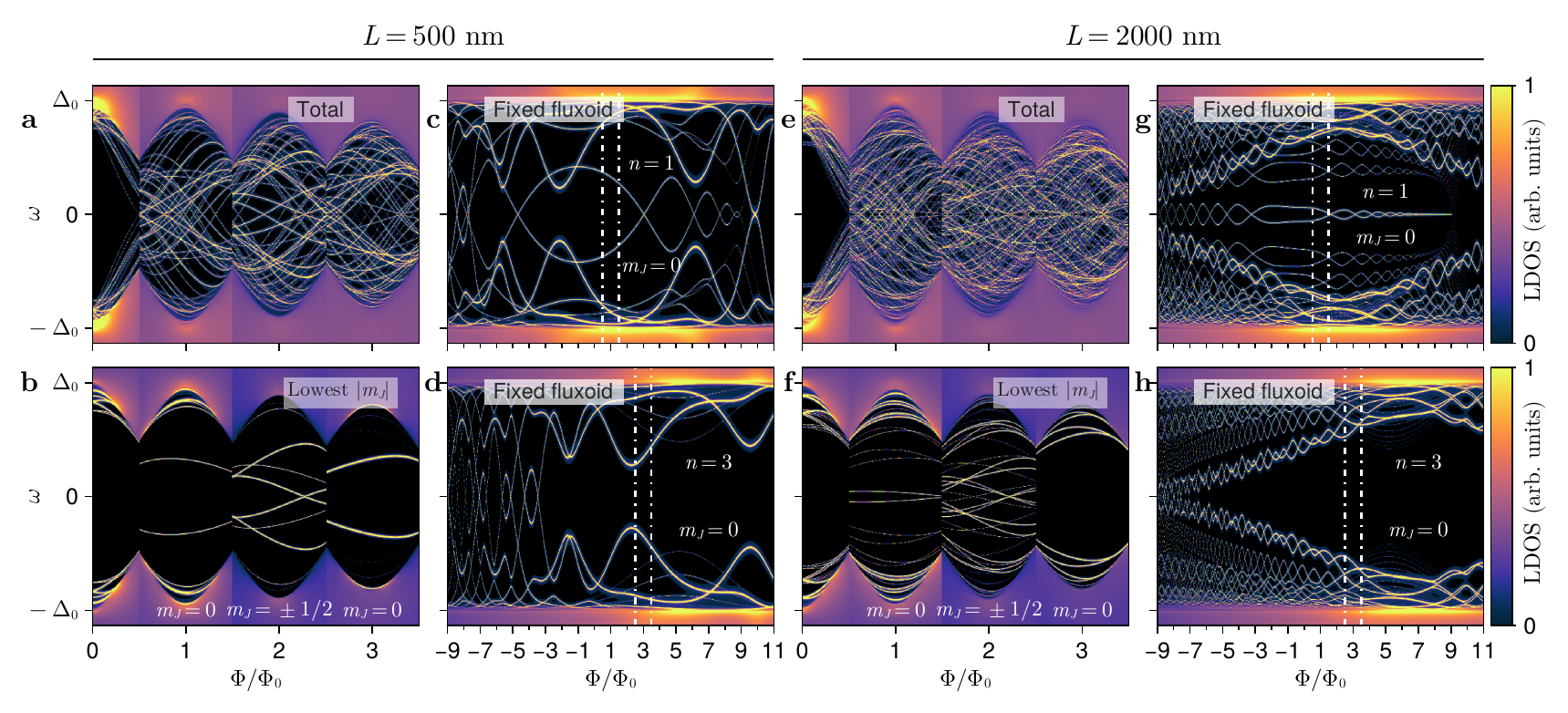}
\caption{\textbf{Solid-core nanowire of finite length.} Same as Fig. \ref{fig:TCMfinite} but for a solid-core nanowire with $d=10$~nm and $R=70$~nm, and nanowire length $L=500$~nm in (a-d) and $L=2000$~nm in (e-h). Other parameters like in Fig. \ref{fig:SCMsemi}.
}
\label{fig:SCMfinite}
\end{figure*}

An equivalent study to Fig. \ref{fig:TCMsemi} but for a semi-infinite solid-core model is shown in Fig. \ref{fig:SCMsemi}. As before, $R=70$~nm and $d=10$~nm, but now we have a dome-like electrostatic potential profile within the semiconductor~\cite{Antipov:PRX18,Mikkelsen:PRX18}, see Fig. \ref{fig:sketch}(b). The chosen parameters correspond to nanowire in the topological regime. Consequently, a Majorana ZEP extends throughout the first lobe corresponding to the second ($m_r=1$) radial momentum subband. There are no Majoranas in the third lobe. In the $n=1$ lobe there is no topological minigap since many CdGM analogs cross zero energy at different fluxes. As thoroughly analyzed in Ref. \cite{Paya:PRB24}, this is a general trait of solid-core nanowires and is an indirect result of the radial profile of the wave function in this model, which may extend closer to the nanowire axis as compared to the tubular-core case. Only in the presence of mode mixing perturbations is it possible to open minigaps (as well as create MBSs in the $n=0,2$ lobes). We ignore this possibility for simplicity, as it does not affect the conclusions about Majorana oscillations studied here.

In Fig. \ref{fig:SCMsemi}(b) we observe the whole Majorana ZEP interval for a fixed $n=1$ fluxoid as before. Note that the $m_r=1$ ZEP that spans the vertical dashed white lines disappears for a magnetic flux of $\Phi/\Phi_0\approx -12$. Subsequently, another ZEP emerges for more negative flux values. This second anomaly corresponds to the $m_r=0$ topological subband.
In the $n=3$ fixed-fluxoid case of Fig. \ref{fig:SCMsemi}(c), there is no ZEP in the third lobe because only at very negative fluxes a $m_r=0$ Majorana ZEP appears. The Majorana localization length is represented above these plots as before, and is systematically dominated by the outer gap. The minimum $\xi_M$ within the first lobe is now more than 3 times larger than in the tubular-core case of Fig. \ref{fig:TCMsemi}.

The finite-length version of solid-core nanowire is analyzed in Fig. \ref{fig:SCMfinite}. Now we consider $L=500$~nm in Fig. \ref{fig:SCMfinite}(a-d) and $L=2000$~nm in Fig. \ref{fig:SCMfinite}(g-h). Once more, even if Majorana oscillations are visible in the artificial fixed-fluxoid panels, these oscillations have always a larger flux period than the flux window of the first lobe. Thus, in practice, Majorana splittings are observed in the LDOS of Figs. \ref{fig:SCMfinite}(b,f), as corresponds to overlapping left and right Majorana wave functions, but full oscillations have no flux range to develop. On the other hand, the total LDOS of Figs. \ref{fig:SCMfinite}(a,e) is covered around zero energy by many longitudinally confined CdGM analog states, the more the larger $L$, which hinders the  observation of the Majorana peaks.

\section{Discussion}
\label{Sec:discussion}

It is possible to understand the absence of Majorana oscillations observed in the previous section analytically for the tubular-core model thanks to the mapping between the single-mode Oreg-Lutchyn Hamiltonian and the $m_J=0$ full-shell nanowire Hamitonian. In Appendix \ref{Ap:oscillations} we derive the Majorana oscillation period $\delta\Phi$, Eq. \eqref{osperiod}, as a function of flux and the rest of the hybrid nanowire parameters. We then compare it with the flux interval that contains Majoranas for a particular lobe $n$ in the topological phase, $I^n_\Phi$, which we can also derive analytically. By inserting numbers for Al/InAs, it is possible to see that $\delta\Phi\gtrsim I^n_\Phi$ and thus Majorana oscillation cannot fully develop in lobes $n=1$ and $n=3$ for the typical range of parameters: $m=0.023m_e$, $\xi_d \in[40, 250]$~nm, $L<1.5~\mathrm{\mu m}$, $R_{\rm LP}\geq R_\mathrm{av}>40$~nm, $\alpha >5$~meVnm and $\Delta_0> 0.05$~meV. A marginal exception to this rule are long ($L>2\mu$m) and narrow ($R_{\rm LP} < 40$~nm) nanowires, which may develop around one oscillation within odd lobes. Obviously, very long wires have the problem that they are more prone to have disorder along the wire and thus Majoranas can be easily destroyed. Even if disorder is sufficiently weak, the amplitude of Majorana oscillations would be exponentially suppressed with length, so that in practice the oscillations would be unobservable. On the other hand, in very narrow nanowires, even though the Majorana oscillation period decreases (with respect to wider ones), the lobe flux window also decreases, again hindering the development of proper Majorana oscillations.

We analyze this narrow regime in Appendix \ref{Ap:DestructiveLP}. There, we consider a hybrid nanowire with $d=10$~nm and $R=30$~nm, both in the tubular-core (Figs. \ref{fig:TCMsemi_des} and \ref{fig:TCMfinite_des}) and the solid-core regimes (Figs. \ref{fig:SCMsemi_des} and \ref{fig:SCMfinite_des}). Keeping $\xi_d$ the same, the full-shell hybrid enters the LP destructive regime when $R_{\rm LP}$ decreases sufficiently~\cite{Schwiete:PRB10}, see Eq. \eqref{eq:desborder}. Two important differences with respect to the non-destructive regime are: (a) there are gapless regions  between lobes and thus the flux window of each lobe decreases as $n$ increases; and (b) the minimum SOC value to enter the topological phase in the tubular-core model strongly decreases with smaller $R$ \cite{Paya:PRB24}. Apart from these differences, the conclusions with respect to the absence of Majorana oscillations are similar to the ones in Sec. \ref{Sec:results}. As long as the hybrid nanowire parameters are within the ranges mentioned above, the oscillation period with flux $\delta\Phi$ is of the order or larger than the Majorana flux interval $I^n_\Phi$, or even the full lobe width.

\section{Conclusions}
\label{Sec:conclusions}

In this work we have analyzed full-shell hybrid nanowires of finite length in the topological regime. On the one hand, subgap CdGM analog states, which in semi-infinite nanowires are Van Hove singularities induced by the superconductor shell on the semiconductor $m_J$ propagating subbands, get longitudinally confined by the finite $L$ and give rise to a dense collection of subgap levels that disperse with flux and that can cross zero energy. On the other hand, Majorana ZEPs that appear for a certain flux interval in the $m_J=0$ sector within odd lobes, split in energy due to the overlap between the left and right MBSs forming at the ends of the finite-$L$ wire.

We have focused on the fate of the Majorana splittings versus magnetic flux. In microscopic numerical simulations performed for Al/InAs hybrids, we have found that while split Majorana ZEPs appear for a wide range of parameters, there is typically no room for proper oscillations within the flux window of actual odd LP lobes. At most, a single parity crossing may occur inside the lobe. This has been confirmed analytically in Appendix \ref{Ap:oscillations}, where we show that in full-shell tubular-core nanowires $\delta\Phi\gtrsim I^n_\Phi$, where $\delta\Phi$ is the Majorana oscillation period and $I^n_\Phi$ is the flux interval that contains Majoranas for a particular lobe $n$. Our prediction is compatible with measurements of the lowest energy mode in full-shell islands using Coulomb spectroscopy \cite{Vaitiekenas:S20}, where finite length-dependent splittings in the first lobe were resolved, but not oscillations.


While our focus in the main text has been the non-destructive LP regime, the corresponding study for a hybrid nanowire in the destructive LP regime can be found in Appendix \ref{Ap:DestructiveLP}. The conclusions are similar in this case. In principle, the Majorana oscillation period $\delta\Phi$ decreases for narrower wires, but so does the lobe width due to the gapless regions that appear between lobes in the destructive LP regime. As a consequence, the flux window available for Majorana oscillations decreases, once more hindering their observation.

We have restricted our study to full-shell nanowires free from all types of disorder or imperfections, since our motivation is to establish the ideal baseline for the  observation of Majorana oscillations. Mode-mixing disorder (i.e., disorder in the transversal direction only such as cross-sectional distortions) was analysed before~\cite{Vaitiekenas:S20,Penaranda:PRR20,Paya:PRB24}. This type of disorder can actually be beneficial for the protection of MBSs, since it can open minigaps around zero energy by coupling different CdGM analog states. Otherwise, this type of disorder would not affect the conclusions about Majorana oscillations analyzed here. Strong disorder along the nanowire axis is however more detrimental~\cite{Pan:PRR20,Das-Sarma:PRB21,Das-Sarma:NP23}. As in conventional Zeeman-driven partial-shell nanowires, we expect this type of disorder to destroy MBSs and/or create quasi-Majoranas~\cite{Kells:PRB12,Prada:PRB12,Rainis:PRB13,Roy:PRB13,Liu:PRB17,Penaranda:PRB18,Vuik:SP19}, which would complicate the interpretation of the subgap phenomenology with local probes~\cite{Prada:NRP20}.
However, full-shell nanowires should be less sensitive than their partial-shell equivalent to some dominant types of disorder, such as trapped charges in the dielectric environment~\cite{Woods:PRA21,Roy:A24}, thanks to the screening of the encapsulating superconductor~\cite{Vaitiekenas:S20}.

Studies of Majorana oscillations, as the one performed in this work, are important because these oscillations are often a target in experiments, since they were predicted to be a characteristic signature of topological Majorana modes~\cite{Das-Sarma:PRB12}. Our results suggest that oscillations versus flux are however not the signature to search for in tunneling spectroscopy experiments on full-shell nanowires when trying to identify potential Majorana peaks. A given, non-oscillating, near-zero peak (or a parity crossing) in a sufficiently short full-shell wire could be interpreted as a mere longitudinally-quantized CdGM analog or a quantum-dot-like state, both topologically trivial. However, the wire could in fact be in the topological phase, and the non-oscillating peak could correspond to a pair of split Majoranas that would be stabilized to zero energy by increasing the length of the device or reducing its disorder. These considerations should also be relevant beyond tunneling spectroscopy experiments, e.g., in Josephson effect devices~\cite{Paya:InPrep}.

All the numerical code used in this manuscript was based on the Quantica.jl package \cite{San-Jose:24a}. The specific code to build the nanowire Hamiltonian and to perform and plot the calculations is available at Ref. \cite{Paya:24a} and Ref. \cite{Paya:24} respectively. Visualizations were made with the Makie.jl package \cite{Danisch:JOSS21}.

\acknowledgments{
This research was supported by Grants PID2021-122769NB-I00, PID2021-125343NB-I00 and PRE2022-101362 funded by MICIU/AEI/10.13039/501100011033, ``ERDF A way of making Europe'' and ``ESF+''.}

\appendix

\section{Hamiltonian}
\label{Ap:Hamiltonian}

A detailed explanation of our models and methodology for studying full-shell hybrid nanowire can be found in Appendix A of Ref. \cite{Paya:PRB24}. There, the reader can find a discussion of the LP effect of the shell, the Bogoliubov-de Gennes (BdG) Hamiltonian, the quantum numbers for a cylindrical full-shell nanowire, the numerical methods for the Green's functions, and the expressions for the LDOS, differential conductance $dI/dV$ and Majorana localization length $\xi_M$. Here, we just provide a summary of the main concepts and the Hamiltonian employed in our calculations to establish the definitions of all  parameters. We refer the reader to Ref. \cite{Paya:PRB24} for further details.

The full-shell nanowire is a hybrid structure consisting of a superconductor shell and a semiconductor core. In the Nambu basis $\Psi=(\psi_\uparrow, \psi_\downarrow, \psi_\downarrow^\dagger, -\psi_\uparrow^\dagger)$, the BdG Hamiltonian is given by
\begin{eqnarray}
H_{\rm{BdG}}=
\begin{bmatrix}
H_0 & \Delta^\dagger\\
\Delta & -\sigma_y H_0^T\sigma_y
\end{bmatrix}.
\label{HBdG}
\end{eqnarray}
Here, $H_0$ is the Hamiltonian of the hybrid system in the normal state and $\Delta$ is the superconducting order parameter, that is only non-zero inside the shell. In the presence of a magnetic field $\vec{B}=B\hat{z}$ applied along the wire's axis, the vector potential in the symmetric gauge reads $\vec{A}=\frac{1}{2}(\vec{B}\times\vec{r})=(-y, x, 0)B_z/2 = A_\varphi\hat{\varphi}$, where $A_\varphi = B r/2$. Here $r$ is the radial coordinate and $\varphi$ denotes the azimuthal angle around $\hat{z}$, as we want to work in cylindrical coordinates. The magnetic field threads a flux through the cylinder, defined as $\Phi = \pi R_\mathrm{LP}^2 B_z$, with $R_\mathrm{LP} =R + d/2$, where $R$ is the semiconductor radius and $d$ is the superconductor thickness. The flux $\Phi$ is thus defined at the mean radius $R_\mathrm{LP}$ of the shell. $\sigma_i$, with $i=(x,y,z)$, are Pauli matrices in the spin sector. We take $\hbar=1$ for simplicity.

The normal Hamiltonian $H_0$ in Eq.~\eqref{HBdG} is composed of $H_{\rm{core}}$ and the shell Hamiltonian in the normal state. In the normal state, the shell is a dense diffusive metal, with much smaller Fermi wavelength than the semiconductor. It is in general quite demanding to include the superconducting shell explicitly in the numerical solution of the Hamiltonian. We then choose to write an \textit{effective} BdG Hamiltonian $H$ of the proximitized nanowire by integrating out the shell degrees of freedom. This procedure introduces a self energy $\Sigma_\mathrm{shell}$ into the Green's function $G(\omega)=\left[\omega - H_\mathrm{core} - \Sigma_\mathrm{shell}(\omega)\right]^{-1}$. This $\Sigma_\mathrm{shell}$ acts on the core surface $r=R$. We thus define the effective BdG Hamiltonian for the system as $H\equiv\omega-G^{-1}(\omega)=H_{\rm{core}}+\Sigma_\mathrm{shell}(\omega)$, which is in general frequency dependent.

It is possible to define a generalized angular momentum as  $J_z = -i\partial_\varphi+\frac{1}{2}\sigma_z +\frac{1}{2}n\tau_z$, which is the sum of the orbital angular momentum $l_z=-i\partial_\varphi$, the spin momentum $s_z=\frac{1}{2}\sigma_z$ and the ``fluxoid momentum" $f_z=\frac{1}{2}n\tau_z$ (given in terms of the fluxoid number $n$), all of them projected along the $z$ direction. For a hybrid wire with cylindrical symmetry, $J_z$ commutes with $H$, $[J_z, H]=0$, so that the eigenvalues $m_J=m_l+m_s+m_n$ of $J_z$ are good quantum numbers of the eigenstates of $H$. Since $m_l\in \mathbb{Z}$ and $m_s\in\pm 1/2$, the possible eigenvalues $m_J$ are~\cite{Vaitiekenas:S20}
\beq
m_J = \left\{\begin{array}{ll}
\mathbb{Z}+ \frac{1}{2} & \textrm{if $n$ is even} \\
\mathbb{Z} & \textrm{if $n$ is odd}
\end{array}\right..
\label{mL}
\eeq
We can thus apply a canonical transformation $\mathcal{U} = e^{-i(m_J-\frac{1}{2}\sigma_z -\frac{1}{2}n\tau_z)\varphi}$ to reduce $H$ to a $\varphi$-independent $4\times 4$ effective Hamiltonian $\tilde H= \mathcal{U}H\mathcal{U}^\dagger$, where
\beqa
\label{eq:solidrot}
\tilde H &=& \left[\frac{p_z^2+p_r^2}{2m}+ U(r)-\mu\right]\sigma_0\tau_z+V_{\rm Z}\sigma_z\tau_0\nonumber\\
&&+\frac{1}{2m r^2}\left(m_J-\frac{1}{2}\sigma_z -\frac{1}{2}n\tau_z+\frac{1}{2}\frac{\Phi}{\Phi_0}\frac{r^2}{R_{\rm{LP}}^2} \tau_z\right)^2\sigma_0\tau_z\nonumber\\
&&-\frac{\alpha(r)}{r}\left(m_J-\frac{1}{2}\sigma_z -\frac{1}{2}n\tau_z+\frac{1}{2}\frac{\Phi}{\Phi_0}\frac{r^2}{R_{\rm{LP}}^2}\tau_z\right)\sigma_z\tau_z\nonumber\\
&&+\alpha(r) k_z\sigma_y\tau_z+\Sigma_\mathrm{shell}(\omega).
\eeqa
Apart from the generalized angular momentum quantum number $m_J$, this Hamiltonian is expressed in terms of the longitudinal momentum operator $p_z=-i\partial_z$, the radial momentum operator $p_r=-i\partial_r$, $p_r^2 = -\frac{1}{r}\partial_r \left(r \partial_r\right) $, the semiconductor effective mass $m$, the chemical potential $\mu$, the flux $\Phi$ and the superconducting flux quantum $\Phi_0=h/2e$.

Even though it is not necessary for the appearance of the topological phase, we also consider the Zeeman effect produced by the magnetic field,
\beq
V_{\rm Z}=\frac{1}{2}g\mu_{\rm B}B_z,
\label{Zeeman}
\eeq
where $\mu_{\rm B}$ is the Bohr magneton and $g$ is the semiconductor Land\'e $g$-factor.

$U(r)$ is the electrostatic potential energy inside the core. This potential is a consequence of the band-bending imposed by the epitaxial core/shell Ohmic contact, which in turn stems from the difference of the Al work function and the InAs electron affinity~\cite{Mikkelsen:PRX18, Liu:PRB21}. We note that the degree of band-bending and precise shape of $U(r)$ depends on the microscopic details of the interface and the self-consistent electrostatic screening. We consider a simple model for $U(r)$ of the form
\beq
\label{pot}
U(r) = U_\mathrm{min} + (U_\mathrm{max}-U_\mathrm{min})\left(\frac{r}{R}\right)^2,
\eeq
for the solid-core nanowire, see Fig.~\ref{fig:sketch}(b).

Concerning the SOC inside the core, we assume it is of the Rashba type and produced by the inversion symmetry-breaking created the core/shell interface. It is thus radial and pointing outwards. For the solid-core nanowire and using a standard approximation from the 8-band model~\cite{Winkler:03}, we can write (see also \cite{Escribano:PRR20})
\begin{eqnarray}
\label{Eq:SOC}
\alpha(r) &=& -\alpha_0\partial_r U(r), \\
\alpha_0 &=& \frac{P^2}{3}\left[\frac{1}{\Delta_{\rm g}^2}-\frac{1}{(\Delta_\mathrm{soff}+\Delta_{\rm g})^2}\right].\nonumber
\end{eqnarray}
Taking the Kane parameter $P = 919.7~\mathrm{meV}\, \mathrm{nm}$, the semiconductor gap $\Delta_{\rm g} = 417$~meV and split-off gap $\Delta_{\rm soff}=390$~meV, relevant for InAs, one obtains $\alpha_0 = 1.19~\textrm{nm}^2$. In our simulations, we take $\alpha_0$ as a free parameter.

The form of the self energy for a diffusive shell in Eq. \eqref{eq:solidrot}, expressed in terms of a decay rate $\Gamma_{\textrm{S}}$ from the core into the shell (in the normal state), has the expression~\cite{Skalski:PR64}
\beq
\label{shelfenergy}
\Sigma_\mathrm{shell}(\omega)=\Gamma_{\textrm{S}}\sigma_0\frac{\tau_x-u(\omega)\tau_0}{\sqrt{1-u(\omega)^2}}.
\eeq
Here, the complex function $u(\omega)$ is obtained from the equation
\beqa
u(\omega)=\frac{\omega}{\Delta(\Lambda)}+\frac{\Lambda}{\Delta(\Lambda)}\frac{u(\omega)}{\sqrt{1-u(\omega)^2}},
\label{Eq:u}
\eeqa
where $\Lambda$ is a pair breaking parameter (introduced by the magnetic field in our case) and the superconductor pairing amplitude $\Delta$ furthermore obeys
\beqa
\ln\frac{\Delta(\Lambda)}{\Delta(0)} &=& -P\left(\frac{\Lambda}{\Delta(\Lambda)}\right),\nonumber\\
P(\lambda\leq 1) &=& \frac{\pi}{4}\lambda,\nonumber\\
P(\lambda\geq 1) &=& \ln\left(\lambda+\sqrt{\lambda^2-1}\right)+\frac{\lambda}{2}\arctan\frac{1}{\sqrt{\lambda^2-1}}\nonumber\\
&&-\frac{\sqrt{\lambda^2-1}}{2\lambda},
\label{LP1}
\eeqa
where $\Delta(0)\equiv \Delta_0$ is the pairing of a ballistic superconductor, i.e., for $\Lambda=0$. Note that $\Lambda$ has energy units and is bounded by $0\leq \Lambda\leq \Delta_0/2$. The equation for $\Delta(\Lambda)$ has to be solved self-consistently.
Equation \eqref{Eq:u} can be rewritten as a fourth-order polynomial with root $u(\omega)$. We choose the solution that leads to the adequate continuity and asymptotic behavior of the retarded Green's functions. As a consequence, $u(\omega\rightarrow 0) \rightarrow 0$.

The superconductor energy gap is given by~\cite{Skalski:PR64}
\beq
\Omega(\Lambda) = \left(\Delta(\Lambda)^{2/3} - \Lambda^{2/3} \right)^{3/2}.
\label{Omega}
\eeq
It closes at $\Lambda = e^{-\pi/4} \Delta_0$, while the pairing becomes zero at $\Lambda = \Delta_0/2$, yielding a gapless superconductor region between these two values of the depairing. Notice that for such region our model is still valid, but for $\Lambda > \Delta_0/2$, $\Delta$ vanishes and the material becomes a metal, with the normal self-energy
\beq
    \Sigma^{\Lambda > \Delta_0/2}_\text{shell}(\omega) = - i \Gamma_{\rm S} \sigma_0 \tau_0.
\eeq

Assuming cylindrical symmetry, a standard Ginzburg-Landau theory of the LP effect \cite{Lopatin:PRL05,Shah:PRB07,Dao:PRB09,Schwiete:PRB10,Sternfeld:PRL11} provides an explicit connection between depairing and flux,
\beqa
\label{depairing}
\Lambda(\Phi) &=& \frac{k_{\rm B} T_{\rm c}\,\xi_{\rm d}^2}{\pi R_{\rm{LP}}^2}\left[4\left(n-\frac{\Phi}{\Phi_0}\right)^2 + \frac{d^2}{R_{\rm{LP}}^2}\left(\frac{\Phi^2}{\Phi_0^2} + \frac{n^2}{3}\right)\right],\nonumber\\
n(\Phi) &=& \lfloor \Phi/\Phi_0\rceil = 0, \pm 1,\pm 2, \dots
\label{LP3}
\eeqa
where $\xi_{\rm d}$ is the diffusive superconducting coherence length and $T_{\rm c}$ is the zero-flux critical temperature. At zero field $\Lambda(0)=0$ and $k_{\rm B} T_{\rm c}\approx\Delta_0/1.76$, where $k_{\rm B}$ is the Boltzmann constant. The way in which $\Lambda(\Phi)$ depends on $\Phi$ gives rise to modulations of the shell gap with flux, defining different regions called LP lobes with fixed fluxoid $n$. The shell gap $\Omega(\Phi)$ is maximum at the center of the lobes, for $\Phi=n\Phi_0$, and it decreases towards the lobe edges. In the non-destructive LP regime the shell gap does not close in between lobes, at half-integer flux $\Phi=(n/2)\Phi_0$. This is the case analyzed in the main text. When the depairing at $\Phi=(n/2)\Phi_0$ is larger than $\Lambda = e^{-\pi/4} \Delta_0$, the shell transitions into the destructive LP regime; this is the case analyzed in Appendix \ref{Ap:DestructiveLP}. The destructive LP regime, defined when a destructive region appears between the $n=0$ and $n=1$ lobes, occurs for
\beq \label{eq:desborder}
C \frac{R_{\rm LP}^2}{\xi_d^2}-\frac{d^2}{4 R_{\rm LP}^2}\leq 1,
\eeq
where $C=\pi e^{-\pi/4}\Delta_0/k_B T_c\approx 2.52$. When $d\rightarrow0$, the destructive regime happens for $R_{\rm LP}/\xi_d\lesssim 0.6$ \cite{Schwiete:PRB10}. 

\subsection{Tubular-core approximation}
\label{Ap:TC}

In the case of a tubular-core nanowire, when the semiconductor thickness $W$ is small as compared to $R$, see Fig. \ref{fig:sketch}(a), we can make some approximations to the Hamiltonian \eqref{eq:solidrot}. For the lowest radial momentum subband, $m_r=0$, it is possible to concentrate all the wave function at an average radius $R_{\rm av}$ within the semiconductor tube section, see Ref. \cite{Paya:PRB24}. We can thus substitute $r=R_{\rm av}$ in Eq. \eqref{eq:solidrot}. Defining $\tilde\mu\equiv\mu-U(R_{\rm av})-\langle p_r^2\rangle/2m$, where $\langle p_r^2\rangle/2m$ represents the radial confinement energy, we can write the tubular-core BdG effective Hamiltonian as

\beqa
\label{TCHam}
\tilde H^{TC} &&= \left(\frac{p_z^2}{2m}-\tilde\mu\right)\sigma_0\tau_z+V_{\rm Z}\sigma_z\tau_0\nonumber\\
&&+\frac{1}{2m R_{\rm av}^2}\left(m_J-\frac{1}{2}\sigma_z -\frac{1}{2}n\tau_z+\frac{1}{2}\frac{\Phi}{\Phi_0}\frac{R_{\rm av}^2}{R_{\rm{LP}}^2} \tau_z\right)^2\sigma_0\tau_z\nonumber\\
&&-\frac{\alpha}{R_{\rm av}}\left(m_J-\frac{1}{2}\sigma_z -\frac{1}{2}n\tau_z+\frac{1}{2}\frac{\Phi}{\Phi_0}\frac{R_{\rm av}^2}{R_{\rm{LP}}^2}\tau_z\right)\sigma_z\tau_z\nonumber\\
&&+\alpha k_z\sigma_y\tau_z+\Sigma_\mathrm{shell}(\omega).
\eeqa
Note that $\tilde\mu$ represents now the Fermi energy (or energy difference between the highest and lowest occupied single-particle states at zero temperature). Moreover, the SOC can be approximated by a constant $\alpha$. The form of the self energy term remains unchanged, but it is now understood to act at $r=R_{\rm av}$ (instead of at $r=R$ like in the solid-core model).
This was dubbed the modified hollow-core approximation of the Hamiltonian in Ref. \cite{Paya:PRB24}.

It is possible \cite{Vaitiekenas:S20} to rewrite Eq. \eqref{TCHam} as
\begin{multline}
\label{TCHamre}
    \tilde{H}^{\rm TC} = \left( \frac{p_z^2}{2 m} - \tilde\mu^{\phi}_{m_J}\right)\sigma_0\tau_z + (V_Z+V_{\rm Z}^{\phi}) \sigma_z + A_{m_J}
    \\ + C_{m_J} \sigma_z \tau_z + \alpha k_z \sigma_y \tau_z + \Sigma_{\text{shell}}(\omega),
\end{multline}
where
\begin{eqnarray}
    \label{LutchynV}
    \tilde\mu^{\phi}_{m_J} &=&  \tilde\mu - \frac{\alpha}{2 R_{\rm av}} - \frac{1}{8 m R_{\rm av}^2} \left( 4 m_J^2 + 1 + \phi^2 \right), \label{Lutchynmu} \\
    V_{\rm Z}^{\phi} &=& \frac{1}{2}\phi \left( \frac{1}{2 m R_{\rm av}^2} + \frac{\alpha}{R_{\rm av}} \right),
    \label{LutchynZeeman}\\
    A_{m_J} &=& -m_J\frac{1}{2 m R_{\rm av}^2}\phi,  \\
    C_{m_J} &=& -m_J \left(\frac{1}{2 m R_{\rm av}^2} + \frac{\alpha}{R_{\rm av}}\right), \label{LutchynC}
\end{eqnarray}
with $\phi=n-\Phi(R_{\rm av})/\Phi_0$. For $m_J=0$, $A_{m_J}=C_{m_J}=0$ and it is then possible to map this Hamiltonian to the conventional single-band Oreg-Lutchyn nanowire model \cite{Lutchyn:PRL10,Oreg:PRL10}. Note that the \textit{effective} Zeeman term $V_{\rm Z}^{\phi}$ has an orbital origin here, and it is present even in the absence of semiconductor $g$ factor. Thanks to this mapping, it is possible to understand that when the wire is threaded by an odd number of fluxoids $n$, the $m_J=0$ subband sector may undergo a topological phase transition and develop MBSs (if the parameters of the wire are such that the topological phase falls within the corresponding lobe).

\begin{figure*}
\centering
\includegraphics[width=\textwidth]{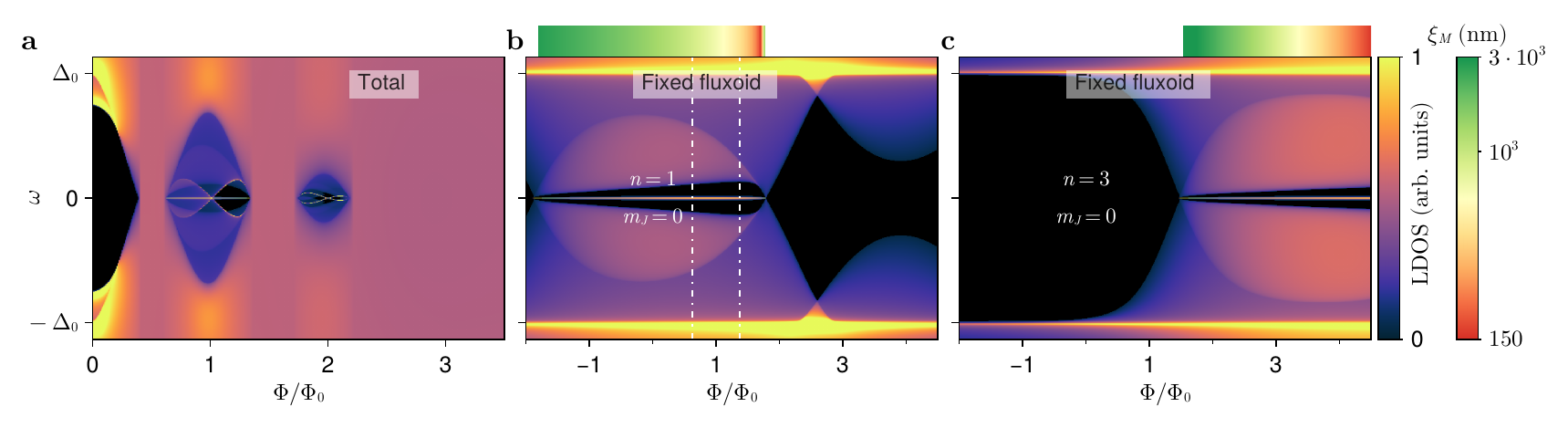}
\caption{\textbf{Tubular-core nanowire of semi-infinite length in the destructive Little-Parks regime.} Same as Fig. \ref{fig:TCMsemi} but for a tubular-core nanowire with $d=10$~nm, $R=30$~nm and $W=10$~nm. The wire is in the destructive LP regime, so that there are gapless regions between lobes and the $n=3$ lobe is closed. Other parameters: $\alpha = 10$~meVnm,  $\mu = 39.1$~meV, $g = 10$,  $\Gamma_{\rm S} = 8 \Delta_0$, $\Delta_0 = 0.23$~meV, $\xi_d = 70$~nm. Minimum $\xi_M$ inside lobe $n=1$: $\sim 450$~nm.
}
\label{fig:TCMsemi_des}
\end{figure*}

\begin{figure*}
\centering
\includegraphics[width=\textwidth]{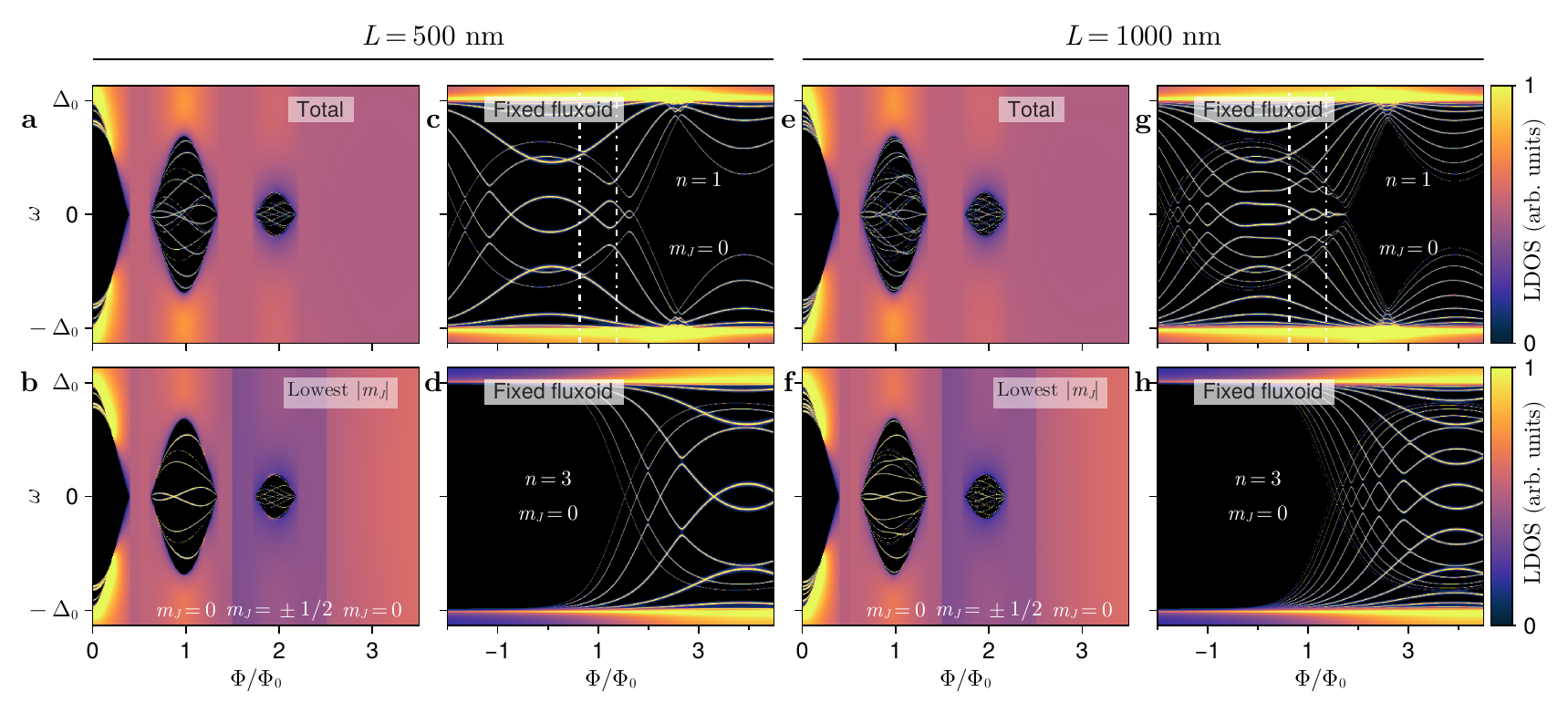}
\caption{\textbf{Tubular-core nanowire of finite length in the destructive Little-Parks regime.} Same as Fig. \ref{fig:TCMfinite} but for a tubular-core nanowire with $d=10$~nm, $R=30$~nm and $W=10$~nm. Other parameters like in Fig. \ref{fig:TCMsemi_des}.
}
\label{fig:TCMfinite_des}
\end{figure*}

\begin{figure*}
\centering
\includegraphics[width=\textwidth]{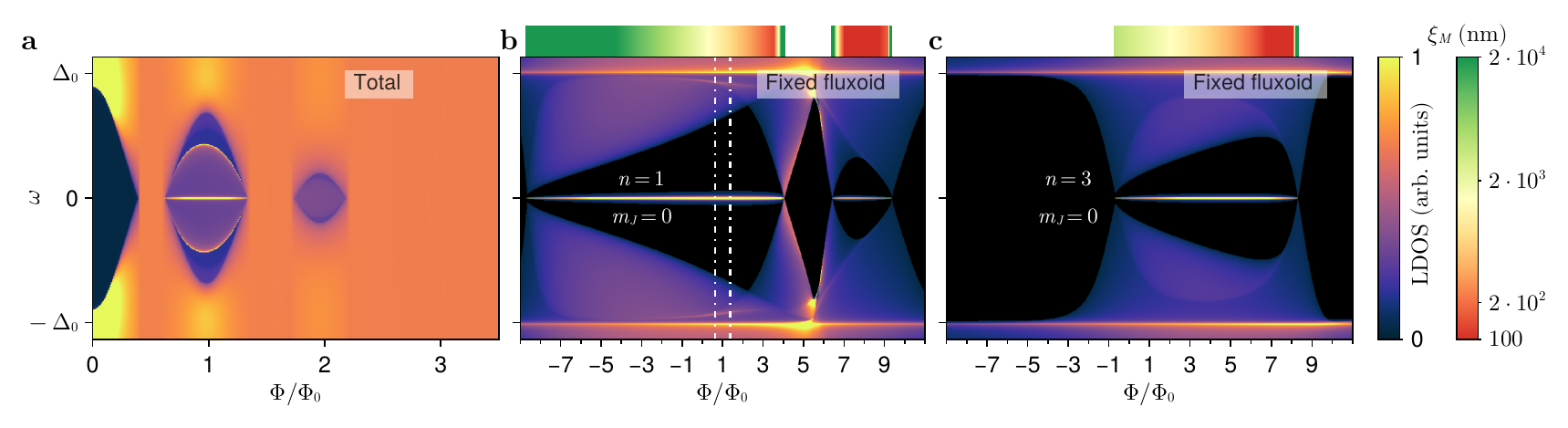}
\caption{\textbf{Solid-core nanowire of semi-infinite length in the destructive Little-Parks regime.} Same as Fig. \ref{fig:TCMsemi} but for a solid-core nanowire with $d=10$~nm and $R=30$~nm. The radial dome-like electrostatic potential profile inside the semiconductor has $U_{\rm min}=-30$~meV and $U_{\rm max}=0$, with $\mu = 2$~meV. Other parameters like in Fig. \ref{fig:TCMsemi_des}, except for $\langle \alpha \rangle = 8.6$~meVnm and $\Gamma_{\rm S} = 40 \Delta_0$. Minimum $\xi_M$ inside lobe $n=1$: $\sim 210$~nm.
}
\label{fig:SCMsemi_des}
\end{figure*}

\begin{figure*}
\centering
\includegraphics[width=\textwidth]{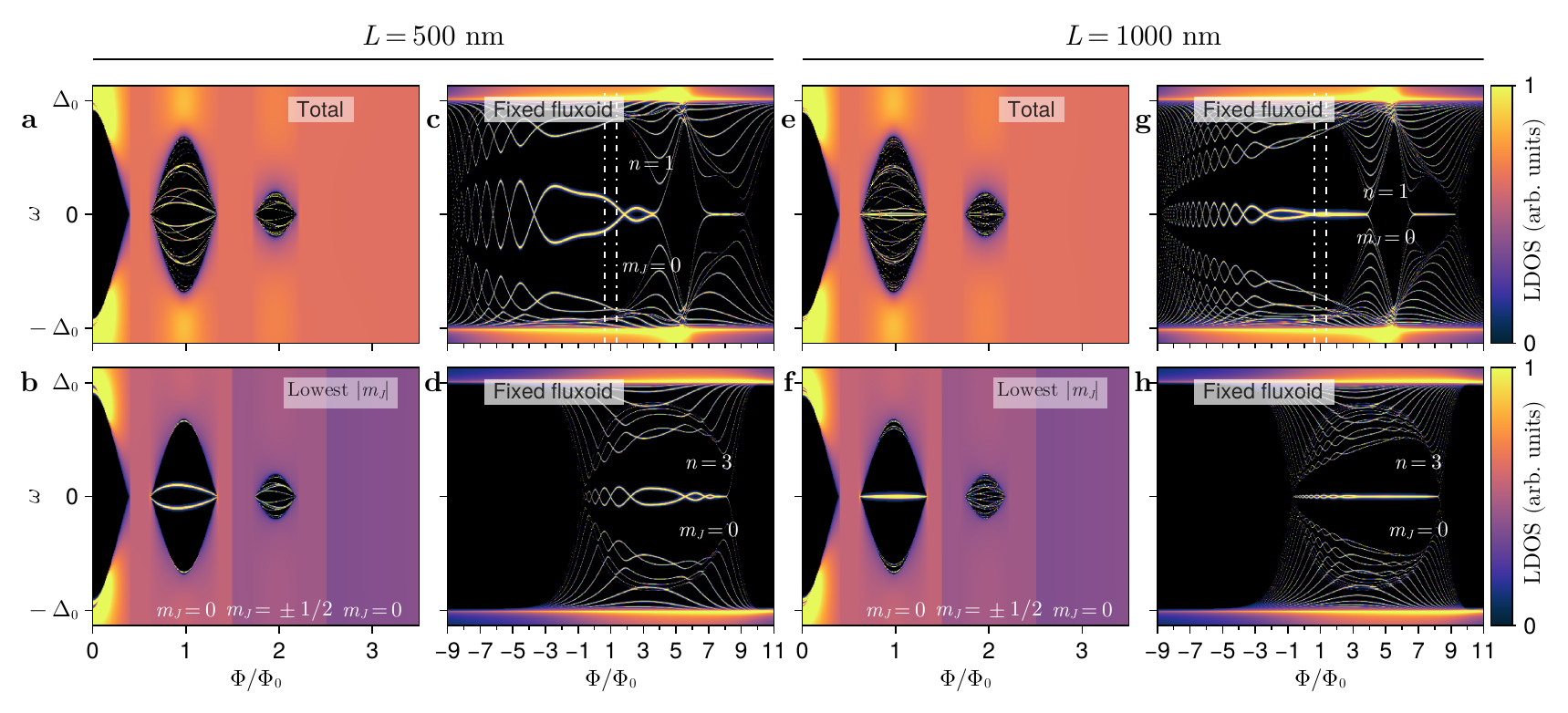}
\caption{\textbf{Solid-core nanowire of finite length in the destructive Little-Parks regime.} Same as Fig. \ref{fig:TCMfinite} but for a solid-core nanowire with $d=10$~nm and $R=30$~nm. Other parameters like in Fig. \ref{fig:SCMsemi_des}.
}
\label{fig:SCMfinite_des}
\end{figure*}

\section{Period of Majorana oscillations}
\label{Ap:oscillations}

In a semi-infinite Majorana nanowire in the topological phase, the wavefunction of the MBS is exponentially localized around the end of the wire as $\sim e^{-z/\xi_M}$, where $\xi_M$ is the Majorana localization length. In Ref. \cite{Das-Sarma:PRB12} it was shown that the exponential decay, however, is also spatially modulated, oscillating as $\sim\cos(k_F z)$, where $k_F$ is the (outer) Fermi wavevector. When the wire has a finite length $L\lesssim\xi_M$, the Majoranas at each end will overlap. As a result, they will hybridize with a typical energy splitting of the form $\sim e^{-2L/\xi_M}\cos(k_FL)$. The dependence of $k_F$ with wire parameters, such as e.g. the chemical potential $\mu$, the Zeeman splitting $V_Z$ or, in full-shell wires, the flux $\Phi$, leads to oscillations of the Majorana splitting as a function of said parameters.

The instantaneous period $\delta x$ of the Majorana oscillations as a function of any parameter $x$ is given by
\beq
\label{period}
\delta x = \left|\frac{2\pi}{L\partial_x k_F(x)}\right|.
\eeq
This expression results from linearizing $k_F(x+\delta x)L \approx k_F(x)L + \delta x\partial_x k_F(x) L$, and equating the last term to a full $2\pi$ phase shift of the cosine.
The above applies equally to an Oreg-Lutchyn nanowire as to a full-shell nanowire. Here we are interested in the latter, where in particular the tuning parameter $x$ is the flux $\Phi$. We will derive the Majorana flux period $\delta\Phi$ in the particular case of the tubular-core model, for which we have the simple analytical mapping to the Oreg-Lutchyn model explained in Appendix \ref{Ap:TC}, and can thus derive a simple form for $k_F(\Phi)$.

We proceed as follows. We first diagonalize the tubular-core Hamiltonian \eqref{TCHamre} but in the absence of the superconductor and for $m_J=0$, i.e., the normal single-band semiconductor Hamiltonian
\beq
\tilde{H}^{\rm TC}_0(k_z)=\frac{k_z^2}{2m} - \tilde\mu^\phi + \alpha k_z\sigma_y+(V_Z+V^\phi_Z)\sigma_z,
\eeq
where $k_z$ is the longitudinal wave vector and
\beq
\label{Lutchynmu0}
\tilde\mu^{\phi} =  \tilde\mu - \frac{\alpha}{2 R_{\rm av}} - \frac{1}{8 m R_{\rm av}^2} \left(1 + \phi^2 \right)
\eeq
is the same as Eq. \eqref{Lutchynmu} but for $m_J=0$. We then equate the eigenvalues to zero to find solutions for $k_z=k_F$. These come in two types: two solutions with high $|k_z|$ (called ``outer'' solution), another two with small $|k_z|$ (``inner''), that actually become zero at the helical transition $|V^\phi_z|=|\tilde\mu^\phi|$. The Majorana oscillations are controlled by the outer solution, which takes the explicit form
\beqa
\label{kF}
k_F &=&\sqrt{2m}\sqrt{m\alpha^2+\tilde\mu^\phi+\epsilon_\phi},\\
\epsilon_\phi  &=&\sqrt{(V_Z+V^\phi_Z)^2+m^2\alpha^4+2m\alpha^2\tilde\mu^\phi}\nonumber.
\eeqa
Some authors have formulated alternative expressions for $k_F$ (in the Zeeman-driven Oreg-Lutchyn model) that remain accurate also for finite $\Delta$ at the expense of making $\alpha = 0$, and viceversa \cite{Das-Sarma:PRB12}.

Inserting Eqs. \eqref{LutchynZeeman} and \eqref{Lutchynmu0} into Eqs. \eqref{kF} and applying Eq. \eqref{period} with $x=\Phi$ yields the Majorana oscillation period with flux in the odd-$n$ LP lobe
\beqa
\label{osperiod}
&&\frac{\delta\Phi}{\Phi_0} = \\
&&\frac{8\pi \epsilon_\phi k_F R_{\rm LP}^2/L}{(\epsilon_\phi\!+\!m\alpha^2)\left[n\!-\!\left(n\!-\!\frac{1}{2}\right)\frac{R_\mathrm{av}^2}{R_{\rm LP}^2}\right]-(V_Z+V_Z^\phi)(1\!+\!2mR_\mathrm{av}\alpha)}.\nonumber
\eeqa
To assess whether at least one Majorana oscillation is visible within an odd lobe, one must compare $\delta\Phi$ to the flux interval $I_\Phi^n = \Phi_R-\Phi_L$  that contains MBSs for lobe $n$. Here $\Phi_L$ and $\Phi_R$ are the minimum and maximum flux with MBSs in the lobe, respectively. A Majorana oscillation will then be visible within the lobe only if $\delta \Phi < I_\Phi^n$. Assuming a Majorana mode is indeed present within the $n$ lobe, and approximating the shell thickness by $d\approx 0$, Eq. \eqref{Omega} yields a simple form for $\Phi_L$ (i.e. the left side of lobe $n$), $\Phi_L\approx \Phi_0\max[n-\frac{1}{2}, n-1.74 R_{\rm LP}/\xi_d]$. $\Phi_R$, the highest flux in the lobe with a Majorana mode, similarly reduces to $\Phi_R = \Phi_0 \min[n+\frac{1}{2}, n+1.74 R_{\rm LP}/\xi_d, \Phi_c/\Phi_0]$. Here $\Phi_c$ is the critical flux, namely the $\Phi$ solution of the equation $\Delta^2+(\tilde\mu^{\phi})^2-(V_Z^\phi)^2=0$.

The condition for full oscillations, $\delta \Phi < I_\Phi^n$, is never satisfied in lobes $n=1$ and $n=3$ for the typical range of parameters of InAs full-shell nanowires ($m=0.023m_e$, $\xi_d \in[40, 250]$~nm, $L<1.5~\mathrm{\mu m}$, $R_{\rm LP}\geq R_\mathrm{av}>40$~nm, $\alpha >5$~meVnm and $\Delta_0> 0.05$~meV). The choice of parameters that maximizes the probability of satisfying the condition and hence of fitting at least one oscillation within the lobe corresponds to long ($L>2\mu$m) and narrow ($R_{\rm LP} < 40$~nm) nanowires. The constraint on $R_{\rm LP}$ is slightly relaxed for higher lobes $n\geq 5$ if $\alpha \lesssim 20$~meVnm, so that oscillations in $R_{\rm LP}<50$nm wires becomes possible, as long as these higher lobes remain open and measurable. This parameter window, however, is quite narrow.



\section{Destructive Little-Parks regime}
\label{Ap:DestructiveLP}

In the main text we have considered a representative case of an Al/InAs full-shell hybrid nanowire in the non-destructive LP regime ($d=10$~nm, $R=70$~nm and $\xi_d=70$~nm). In this Appendix we analyze a representative case in the destructive LP regime, a narrower full-shell hybrid nanowire (with $R=30$~nm but with the same shell thickness $d=10$~nm and coherence length $\xi_d=70$~nm).

\subsection{Tubular-core model}
\label{Ap:TCM}

We first consider the tubular-core model for the full-shell nanowire, with a semiconductor tube thickness $W=10$~nm. In analogy to Fig. \ref{fig:TCMsemi} of the main text, in Fig. \ref{fig:TCMsemi_des}(a) we plot the LDOS at the end of a semi-infinite nanowire as a function of normalized flux. Since the wire is in the destructive LP regime, now there are gapless regions between LP lobes for fluxes around half-integer multiples of the superconducting flux quantum $\Phi_0$. Note that the maximum lobe height decreases with the lobe number $n$ (as corresponds to a wire with finite $d$), as well as the lobe flux window. For these parameters, the $n=3$ lobe (and subsequent ones) is already closed. The wire is in the topological regime and a Majorana ZEP can be seen crossing the whole $n=1$ lobe. Additionally, low lying CdGM analog states disperse with flux and cross zero energy in both the $n=1,2$ lobes. A topological minigap at the right end of the first LP lobe can be observed. The $m_J=0$, $n=1$ fixed-fluxoid simulation analogous to Fig. \ref{fig:TCMsemi}(b) is presented in Fig. \ref{fig:TCMsemi_des}(b). We can observe that the Majorana localization length $\xi_M$ (depicted above the panel) is dominated by the outer ($k_z=k_F$) $m_J=0$ gap. (A similar simulation for $n=3$ is presented in Fig. \ref{fig:TCMsemi_des}(c), although in this case it is irrelevant since the third lobe is closed).

The case of a finite-length tubular-core hybrid nanowire is analyzed in Fig. \ref{fig:TCMfinite_des}. For $L=500$~nm, the $m_J=0$ contribution to the LDOS in the first lobe, Fig. \ref{fig:TCMfinite_des}(b), shows what appears to be a couple of Majorana oscillations versus flux. By checking the $n=1$ fixed-fluxoid plot of Fig. \ref{fig:TCMfinite_des}(c), we can understand that it is just a single Majorana parity crossing. The split levels go to zero energy at the lobe edges in Fig.\ref{fig:TCMfinite_des}(b) as a consequence of the LP parent gap closing. Thus, in this case there are no complete Majorana oscillations within the lobe either. Incidentally, we note that in Fig. \ref{fig:TCMfinite_des}(b) there are color discontinuities at half integer $\Phi/\Phi_0$ values. These are non physical, just artifacts of selecting the lowest $|m_J|$ contribution to the total LDOS. The complete sum, Fig. \ref{fig:TCMfinite_des}(a), is continuous (although not its first derivative with respect to flux). A similar study for $L=1000$~nm is shown on the right half of Fig. \ref{fig:TCMfinite_des}. Although one could interpret that there are a couple of oscillations in  the first lobe of Fig. \ref{fig:TCMfinite_des}(f), in this case there is only one true complete Majorana oscillation according to Fig. \ref{fig:TCMfinite_des}(g).

The possible presence of Majorana oscillations in the first lobe of Figs. \ref{fig:TCMfinite_des}(a,e) is accompanied, and thus masked, by the presence of zero-energy crossings and apparent oscillations of other subgap levels. Again, these apparent oscillations come from a combination of the dispersion with flux of the CdGM levels and the parent-gap closing at the lobe edges. In principle, one could distinguish between Majoranas and CdGM states in $dI/dV$ measurements because the Majorana levels are states bound to the hybrid wire end, thus more strongly coupled to the tunnel-contact probe, whereas the CdGM states are delocalized along the length of the wire. This should translate into different total peak weight in tunneling spectroscopy. However, the Majorana localization length $\xi_M$ and the nanowire length $L$ in realistic experiments are often not that different, so the peak weight difference may be small.

We emphasize that the hybrid wire analyzed in this section is a borderline case where the Majorana oscillation period is comparable to or slightly smaller than the flux interval that contains Majoranas for the $n=1$ lobe, $\delta\Phi\sim I^{n=1}_\Phi$. In general, for other (realistic) parameters, we find $\delta\Phi\gtrsim I^{n}_\Phi$ as discussed in the main text and Appendix \ref{Ap:oscillations}.

\subsection{Solid-core model}
\label{Ap:SCM}

Finally, we consider a solid-core full-shell hybrid nanowire in the destructive LP regime. The case of a semi-infinite wire can be seen in Fig. \ref{fig:SCMsemi_des}. As in the tubular-core case of Fig. \ref{fig:TCMsemi_des}(a), there are gapless flux intervals between lobes and the third and subsequent LP lobes have disappeared altogether. A Majorana ZEP is also present across the whole $n=1$ lobe, but now there is no topological minigap at any flux. As in the solid-core case of Fig. \ref{fig:SCMsemi} in the main text, the Majorana zero mode coexists with a LDOS background coming from the rest of the occupied CdGM analogs. This is the typical behaviour of solid-core full-shell nanowires. The Majorana localization length is $\xi_M\approx 210$~nm for the parameters selected in this figure.

For a $L=500$~nm hybrid nanowire we find that the Majorana modes split in energy, as corresponds to overlapping left- and right-end MBSs, see Fig. \ref{fig:SCMfinite_des}(b), but there is not a single Majorana oscillation inside the lobe, as can be checked in Fig. \ref{fig:SCMfinite_des}(c). The closing of the split peaks at the lobe edges in Fig. \ref{fig:SCMfinite_des}(b) is a consequence of the parent-gap closing at the lobe edges in the destructive LP regime. In this case, the CdGM levels do not cross zero energy and are in general at higher energies than the Majorana states, as can be seen in Fig. \ref{fig:SCMfinite_des}(a). For a longer hybrid nanowire, there could be in principle more chances to observe Majorana oscillations within the first lobe. However, this is not the case, see Fig. \ref{fig:SCMfinite_des}(e). The Majorana splitting is so small, see Fig. \ref{fig:SCMfinite_des}(f), that it could be mistaken with a true Majorana zero mode. In this case, only the longitudinally-confined CdGM levels reveal that the wire has indeed a finite length.

\bibliography{biblio}

\end{document}